# ACCRETION PROCESSES

ALESSANDRO MORBIDELLI
CNRS, OBSERVATOIRE DE LA COTE D'AZUR, NICE, FRANCE

## Abstract

In planetary science, accretion is the process in which solids agglomerate to form larger and larger objects and eventually planets are produced. The initial conditions are a disc of gas and microscopic solid particles, with a total mass of about 1% of the gas mass. These discs are routinely detected around young stars and are now imaged with the new generation of instruments. Accretion has to be effective and fast. Effective, because the original total mass in solids in the solar protoplanetary disk was probably of the order of ~ 300 Earth masses, and the mass incorporated into the planets is ~100 Earth masses. Fast, because the cores of the giant planets had to grow to tens of Earth masses in order to capture massive doses of hydrogen and helium from the disc before the dispersal of the latter, i.e. in a few millions of years.

The surveys for extrasolar planets have shown that most stars have planets around them. Accretion is therefore not an oddity of the Solar System. However, the final planetary systems are very different from each other, and typically very different from the Solar System. Observations have shown that more than 50% of the stars have planets that don't have analogues in the Solar System. Therefore the Solar System is not the typical specimen. Thus, models of planet accretion have to explain not only how planets form but also why the outcomes of the accretion history can be so diverse.

There is probably not one accretion process but several, depending on the scale at which accretion operates. A first process is the sticking of microscopic dust into larger grains and pebbles. A second process is the formation of an intermediate class of objects called planetesimals. There are still planetesimals left in the Solar System. They are the asteroids orbiting between the orbits of Mars and Jupiter, the trans-Neptunian objects in the distant system and other objects trapped along the orbits of the planets (Trojans) or around the giant planets themselves (irregular satellites). The Oort cloud, source of the long period comets, is also made of planetesimals ejected from the region of formation of the giant planets. A third accretion process has to lead from planetesimals to planets. Actually, several processes can be involved in this step, from collisional coagulation among planetesimals to the accretion of small particles under the effect of gas drag, to giant impacts between protoplanets. This chapter will detail all these processes, adopting a historical perspective: i.e. from the classic processes investigated in the past decades to those unveiled in the last years.

The quest for planet formation is still ongoing. Open issues remain and exciting future developments are expected.

Keywords: dust, grains, pebbles, planetesimals, planets, collisions, coagulation, gas-drag, streaming instability, turbulence.

---------------------------------------------------------------------------------

The gas that forms the star and the circumstellar disc around it, also called *protoplanetary disk*, is made mostly of hydrogen (H) and helium (He), with a small fraction (of the order of 1% in mass) of heavier elements. Some of these elements are inherited from the interstellar medium in the form of small grains, still visible today in meteorites as nanometre- to micrometre-scale particles, rich in carbon and silicates (such as graphites, nanodiamonds, silicon-carbides). Others are in vapour form and they condense in the disc when the temperature of the gas decreases enough, also forming micron-size particles. The observation of protoplanetary discs around young stars reveals that discs are dusty; micrometer grains make most of the gas opacity.

Therefore, planet formation is a spectacularly successful accretion story, in which solids grow by 13

orders of magnitude in size, passing from micrometres to tens of thousands of kilometres in diameter. How this happens is not yet fully clear. Different processes are at work at different scales. This chapter reviews the historical attempts to understand how planet formation works, their problems and the modern ideas that look the most promising at the current time. The first part addresses the formation of *planetesimals*, bodies of sizes typical of asteroids, comets and trans-Neptunian objects, which clearly constitute a fundamental stage in the growth to planet-size bodies. The second part discusses how planetesimals can become planets. Different kinds of planets will be described: terrestrial, super-Earths and giants, with different compositions and presumably different accretion histories.

## Models of Planetesimal Formation

Observations show that young stars are surrounded by disks of gas and dust particles. The formation of the disk, first predicted by Kant and Laplace, is a natural consequence of conservation of angular momentum during the contraction phase of the gas towards the central star.

The evolution of a particle in a disk predominantly made of gas is dictated by the particle's stopping time $T_s$, which is the time needed to erase any differential velocity of the particle relative to the gas due to collisions with the gas molecules. For small particles (such that the mean free path of gas molecules is larger than about half of the particle size), the Epstein gas drag gives $T_s$ proportional to the particle size, while for bigger particles, the Stokes' drag gives $T_s$ proportional to the square of the size.

To describe a particle in orbiting in a disk, cylindrical coordinates are the natural choice. The stellar gravitational force can be decomposed into a radial and a vertical component. The radial component is cancelled by the centrifugal force. The vertical component, $F_{g,z}= -m \Omega^2 z$, where m is the mass of the particle and z its vertical coordinate, instead accelerates the particle towards the midplane, until its velocity $v_{settle}$ is such that the gas drag force $F_D=mv_{settle}/T_s$ cancels $F_{g,z}$. This sets $v_{settle}=\Omega^2 z\, T_s$ and gives a settling time $T_{settle}=z/v_{settle}=1/(\Omega^2 T_s)$.

If the drag is in the Epstein regime, this gives

$T_{settle} \sim \Sigma_g/(\rho_p \Omega\, r) \sim (\rho_g c_s)/(\rho_p \Omega^2 r)$

where $\Sigma_g$ is the disc's surface density, $\rho_g$ its volume density, $c_s$ is the sound speed (of the order of the thermal velocity dispersion of gas molecules), $\rho_p$ the bulk density of the particles, r their physical radius, and $\Omega$ is the local orbital frequency of the gas (Goldreich and Ward, 1973; Weidenschilling, 1980). Assuming Hayashi's (1981) disk model, where $\Sigma_g/\Omega$ is independent of the distance from the central star, the sedimentation time depends only on particle's size and is :

$T_{sed} \sim 10^3/r$ yr

where r is given in cm. This timescale, however, is computed assuming a quiet, laminar nebula. If the nebula is turbulent the sedimentation time can become much longer; moreover, the particles cannot sediment indefinitely but have to form a layer around the midplane with some finite thickness.

Assuming negligible turbulence and unlimited sedimentation, Goldreich and Ward (1973) pointed out that eventually the density of solids has to become comparable to, or larger than, the density of gas. In the ultra-dense layer of solids, particle collisions would damp the mutual relative velocities. Thus, the disk of solids would become unstable against local density perturbations under the action

of its own gravity. This would give rise to clumps of solid material of size-scale

$$\lambda = 4\pi^2 G \sigma_p / \Omega^2,$$

where G is the gravitational constant and $\sigma_p$ is the surface density of the particle layer. Each clump would contract under its own gravity (Goldreich and Ward, 1973). This would produce planetesimals of about 10 km in size at 1 AU, assuming typical values for $\sigma_p$ and for the planetesimals' bulk densities.

However, Weidenschilling (1980) demonstrated that this scenario faces an internal inconsistency. In fact, the gas, being sustained by its own pressure, behaves as if it felt a central star of lower mass, and therefore rotates more slowly than a purely Keplerian orbit at the same heliocentric distance. Solid particles tend to be on Keplerian orbits and therefore have a larger speed than the gas. If sedimentation really occurred, the solids -once reaching a large enough density- would entrain the gas near the mid-plane to a quasi-Keplerian speed. At higher altitudes from the mid-plane, however, the gas would still rotate around the Sun with a sub-Keplerian speed. The friction between the two layers of gas moving at different velocities would generate turbulence, through the Kelvin-Helmoltz instability mechanism. But turbulence would inhibit the settling of the solid particles to the mid-plane, in first place. In conclusion, particles cannot sediment into an ultra-dense layer, and therefore the conditions for the development of gravitational instabilities cannot be fulfilled.

If the Goldreich-Ward scenario has to be rejected, then one has to envision that the formation of planetesimals occurs by pair-wise collision and sticking of ever bigger grains. However, this growth mode leads to a problem, known as the *metre-size barrier*. This barrier problem has two aspects, both related to gas-drag, as detailed below.

As said above, the gas orbits the star slightly more slowly than the Keplerian speed. It's orbital velocity is $v_{orb}=(1-\eta)v_K$, where is the $v_K$ Keplerian velocity and $\eta$ is a parameter dependent on the radial pressure gradient (typically of order 0.002-0.003). The solids, instead, in absence of gas-friction, would orbit at Keplerian speed. Because of the difference in orbital speed between the gas and the solids, the latter "see" a headwind with a velocity $\eta v_K$, which results in a drag force $F_D$, with its characteristic stopping time $T_s$. This time has to be compared with the characteristic Keplerian time $T_K = 1/\Omega$. So, it is convenient to introduce the Stokes' number $\tau_s = T_s/T_K$

If $\tau_s \gg 1$, the particle is almost decoupled from the gas. If $\tau_s \ll 1$, it is strongly coupled to the gas, and it tends move with the gas flow. In both cases the deceleration of the particle motion is very small, so the particles tend to stay on closed orbits. But for intermediate values of $\tau_s$, the deceleration of the particle's orbital motion due to gas-drag cannot be neglected. Like all decelerations in orbital mechanics, it causes a radial drift of the particles towards the central star, namely an inward-spiralling trajectory. The maximum effect occurs when $\tau_s \sim 1$, which occurs for ~30cm at 1 AU in the Stokes regime (or 10cm at 10 AU, or 1cm at 50 AU, both in the Epstein regime). For the Hayashi MMSN model, at 1 AU the particle's maximal radial velocity induced by the gas drag is then of order 100m/s (Weidenschilling, 1977). So $\tau_s \sim 1$ particles should fall onto the Sun from 1 AU in about 100 years. This is less than the estimated collisional growth time, so that these objects should fall onto the Sun before they can grow massive enough to decouple themselves from the gas. This is the first aspect of the metre-size barrier problem.

The second aspect is that particles of different sizes spiral inwards with different radial velocities, scaling as $v_r \sim \tau_s \eta v_K$. This leads to mutual collisions, with relative velocities essentially determined by the difference in radial drift speeds. In the inner disc, these velocities can easily exceed several metres per second. Using these velocities as input in a model calibrated on laboratory experiments of the outcome of pair-wise collisions between particles of different mass ratios, Windmark et al.

(2012) found that in the inner part of the disk particles can difficultly grow beyond a millimetre in size. A bouncing barrier prevents particles to grow beyond this limit (Fig 1). If a particle, for some lucky reason, managed to grow to ~10cm, its growth could potentially resume by accreting tiny particles. But, as soon as particles of comparable sizes hit each other, erosion or catastrophic fragmentation occurs, thus preventing the formation of planetesimal-size objects.

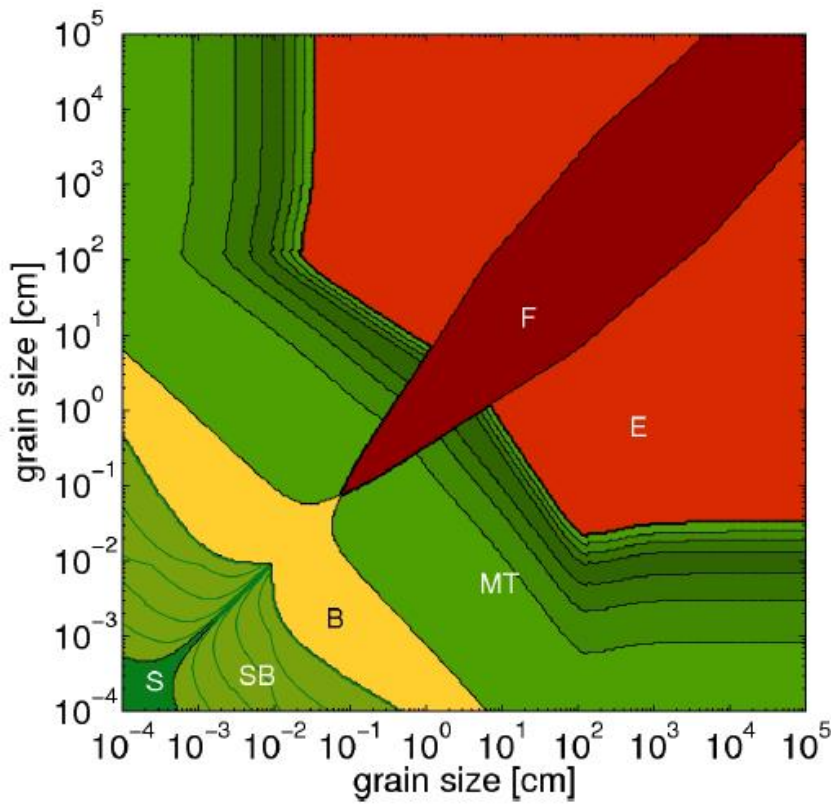

( Windmark et al. 2012 )

Fig. 1. Map of collisional outcome in the disk. The sizes of colliding particles are reported on the axes. The colours denote the result of each pair-wise collision. Green denotes growth, red denotes erosion and yellow denotes neither of the above (i.e. a bounce). The label S stands for sticking, SB for stick and bounce, B for bounce, MT for mass transfer, E for erosion and F for fragmentation. Form Windmark et al. (2012). This map is computed for compact (silicate) particles, at 3 AU.

In the outer part of the disk, due to the lower velocities and the sticking effect of water ice, particles can grow to larger sizes, but never significantly exceeding centimetre-size (Birnstiel et al., 2016). However, it has been proposed that if particles, instead of compact, are very porous, they could absorb better the collisional energy, thus continuing to grow without bouncing or breaking (Okuzumi et al. 2012). Very porous planetesimals could be formed this way. Their low density would make them drift in the disk very slowly, thus circumventing the drift-aspect of the metre-size barrier mentioned above. Eventually, these planetesimals would become compact under the effect of their own gravity and of the ram pressure of the flowing gas (Kataoka et al., 2013). This formation mechanism for planetesimals is still not generally accepted in the community. At best, it could work only in the outer part of the disk, where icy monomers have the tendency to form very porous structures, but not in the inner part of the disk, dominated by silicate particles. Moreover, meteorites show the interior structure of asteroids is made mostly of compact particles of 100 microns to a millimetre in size, called *chondrules*, which is not consistent with the porous formation mode.

To avoid all these bottlenecks in the growth process, the model of planetesimal formation via gravitational instability has been recently resurrected, although in a more complex form with respect to the original Goldreich-Ward approach. It has been showed that, if there is turbulence in the disk, either generated by the Kelvin-Helmholtz instability or other mechanisms, particles can be clumped into turbulent structures (vortices or low-vorticity regions, depending on their Stokes' number; Johansen et al., 2007; Cuzzi et al., 2008). If the disk is not turbulent, particles can still clump due to the so-called *streaming instability* (Youding and Goodman, 2005). Although originally discovered as a linear instability (see Jacquet et al., 2011), this instability raises even more powerful effects which can be qualitatively explained as follows. This instability arises from the aforementioned speed difference between gas and solid particles. If this differential speed causes gas-drag onto the particles, the friction exerted from the particles back onto the gas accelerates the gas and diminishes its difference from the Keplerian speed. Thus, if there is a small overdensity of particles, the local gas is in a less sub-Keplerian rotation than elsewhere; this in turn reduces the local headwind on the particles, which therefore drift more slowly towards the star. Consequently, an isolated particle located farther away in the disc, feeling a stronger headwind and drifting faster towards the star, eventually joins this overdensity region. This enhances the local density of particles and reduces further its radial drift. It's easy to see that this process drives a positive feedback, i.e. and instability, where the local density of particles increases exponentially with time.

The particle clumps generated by disc turbulent and/or the streaming instability, if dense enough, can become self-gravitating and contract to form planetesimals. Numerical simulations of the streaming instability process (Johansen et al., 2015; Simon et al., 2016) show that planetesimals of a variety of sizes can be produced, but those that carry most of the final total mass are those of ~100km in size. The size of 100km is indeed prominent in the observed size-frequency distributions of both asteroids and Kuiper-belt objects. Thus, in these models suggest that planetesimals form (at least preferentially) big, in stark contrast with the collisional coagulation model in which planetesimals would grow progressively from pair wise collisions. If the amount of solid mass in small particles is large enough, even Ceres-size planetesimals can be directly produced from particle clumps (Fig. 2).

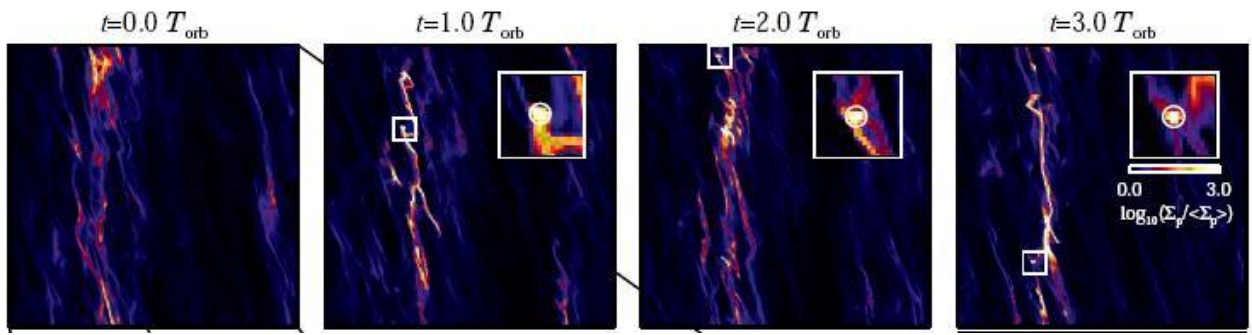

Fig. 2: Concentration of solid particles in a turbulent disk and onset of gravitational instability (from Johansen et al., 2007). The x and y-axes represent respectively the radial and the azimuthal directions in the disk. The colours illustrate the local density of solids, relative to their mean density (black stands for no concentration, yellow for high concentration). The time in orbital periods is reported above each panel. Notice the progressive local concentration of the solid particles with time. The box inside the panels magnifies the density structure in the vicinity of the most massive self-gravitating clump of solids, whose total mass eventually exceeds the mass of Ceres.

A concern about these large-scale particle concentration models is that typically very large particles are needed for optimal concentration (at least decimetre in size when the models are applied to the asteroid belt). Chondrules of typical sizes from 0.1 to 1 mm are ubiquitous in primitive meteorites, but such small particles are very hard to concentrate in vortices or through the streaming instability. Recent high-resolution numerical simulations (Yang et al., 2016) show that chondrule-size particles can trigger the streaming instability only if the initial mass ratio between these particles and the gas is larger than 4% (Fig. 3). However, the initial solid/gas ratio of the Solar System disk is estimated to have been 1%. Thus, at face value, planetesimals should have not formed as agglomerates of chondrules. A possibility is that future simulations with even-higher resolution and run on longer timescales will show that the instability can occur for a smaller solid/gas ratio, approaching the value measured in the Sun.

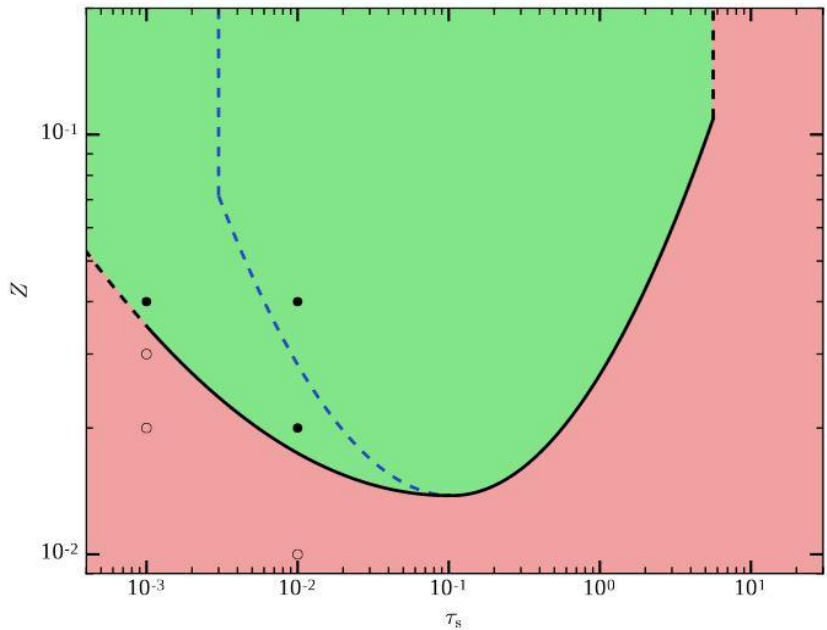

Fig. 3. The green region shows the solid/gas mass ratio Z required for particles of Stokes number $\tau_s$ to generate the streaming instability (from Yang et al., 2016). If Z is not large enough, the instability cannot occur (red region). The blue dashed line shows the limit for streaming instability found in an earlier work using simulations with more limited resolution. Thus, there is some hope that future simulations, with even larger resolution, might show that the actual Z-threshold for the streaming instability is below 1%, the Solar solid/gas ratio.

Another possibility, that is currently actively investigated, is that drifting particles first accumulate at distinct radii in the disc where their radial speed is slowest and then, thanks to the locally enhanced particle/gas ratio, trigger locally the streaming instability. Two locations have been identified for this preliminary radial pile-up. One is in the vicinity of the so-called *snowline*, i.e. the location where water transitions from vapour to solid form (Ida and Guillot, 2016; Schoonenberg and Ormel, 2017). The other is in the vicinity of 1 AU (Drazkowska et al., 2016). Thus, these would be the two locations where planetesimals could form very early in the proto-planetary disk. Elsewhere, the conditions for the streaming instability would not have met in an early disk. In these regions, planetesimals could have formed only later on, when gas was substantially depleted by photo-evaporation from the central star, provided that the solids remained abundant (Throop and Bally, 2005; Carrera et al., 2017).

At least at the qualitative level, this picture is consistent with available data for the Solar System. The meteorite record reveals that some planetesimals formed very early (in the first few $10^5$y). Because of the large abundance of short-lived radioactive elements present at the early time, these first planetesimals melted and differentiated, and are today the parent bodies of iron meteorites. But a second population of planetesimals formed 2 to 4 My later. These planetesimals did not melt and are the parent bodies of the primitive meteorites called the *chondrites*. It is suggestive to speculate that differentiated planetesimals formed at the 1 AU pile-up location advocated by Drazkowska et al., whereas the undifferentiated planetesimals formed beyond this location, in the asteroid belt. On the other hand, Jupiter should also have formed very quickly, in order to capture a large amount of gas from the disc into its atmosphere before disc removal. Recent cosmochemical constraints suggest that Jupiter reached 20 Earth masses in less than one million years (Kruijer et al., 2017). If Jupiter formed so early, its precursor planetesimals should have formed even earlier and this could have occurred at the snowline location. On the other hand, the low bulk densities suggest that most trans-Neptunian objects smaller than D=350 km (Brown, 2013) and comets are undifferentiated, which suggests that, farther from the snowline, early accretion could not be possible, like in the asteroid belt.

## Making PLANETS

Planetesimals somehow formed, via one of the processes described before. The term 'planetesimals' here means objects large enough to avoid significant drifting towards the central star by gas drag. This definition requires that planetesimals have a size of at least a kilometre. It is possible that the planetesimals are much larger than this size, if they are formed by self-graviting clumps, as discussed above.

Once planetesimals appear in the disk, accretion continues by mutual coagulation between planetesimals, through low-velocity two-body collisions. Gravity plays an important role; it bends the trajectories of the colliding objects, which effectively increases the collisional cross-section by a factor

$$F_g = 1 + V_{esc}^2/V_{rel}^2,$$

where $V_{esc}$ is the mutual escape velocity defined as $V_{esc} = [2G (M_1+M_2)/(r_1+r_2)]^{1/2}$, $M_1$, $M_2$, $r_1$, $r_2$ are the masses and radii of the colliding bodies, $V_{rel}$ is their relative velocity before the encounter and G is the gravitational constant. $F_g$ is called the *gravitational focussing factor* (Fig. 4).

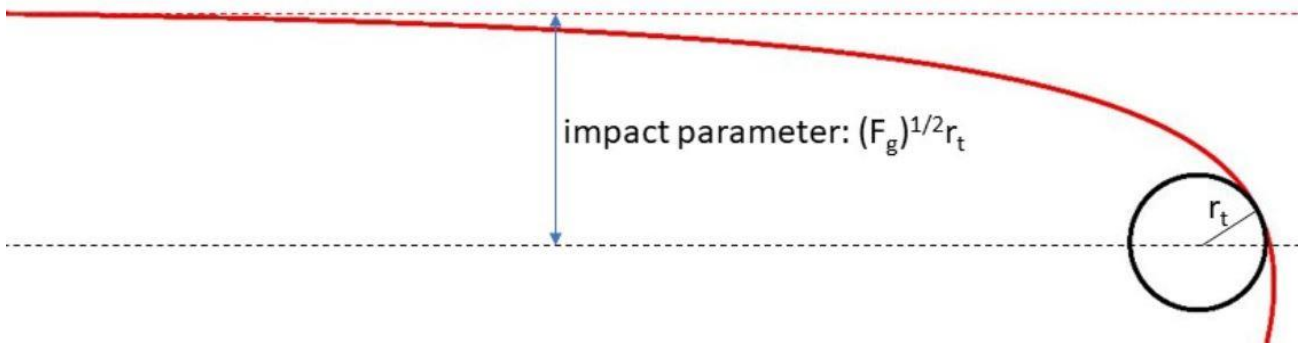

Fig. 4. A schematic illustration of how gravity increases the collisional cross-section by bending the trajectories of the colliding body. Here $r_t$ is the radius of the target (back circle). The dashed red line is the asymptotic trajectory of the projectile and the solid red curve is its real hyperbolic trajectory, bended by the gravity of the target.

Thus, the mass accretion rate of an object is

$$dM/dt \sim r^2 F_g \sim M^{2/3} F_g$$

where the bulk density of planetesimals is assumed to be independent of their mass, so that the planetesimal physical radius r is proportional to $M^{1/3}$. This formula can lead to two very distinct growth modes:

**i) Runaway growth**

Imagine the situation in which one body, of mass $M_1$, has an escape velocity $V_{esc(1)}$ much larger than its relative velocity $V_{rel}$ with respect to the rest of the planetesimal population. Then one can approximate $F_g$ with $V_{esc(1)}^2/V_{rel}^2$. Notice that the approximation $r \sim M^{1/3}$ makes $V_{esc(1)}^2 \sim M_1^{2/3}$. Substituting this expression into the mass-growth equation above, leads to:

$$dM_1/dt \sim M_1^{4/3}/V_{rel}^2,$$

or

$$1/M_1 \, dM_1/dt \sim M_1^{1/3}/V_{rel}^2$$

Consider now a second body of mass $M_2 < M_1$ for which the same approximations hold, so that

$$1/M_2 \, dM_2/dt \sim M_2^{1/3}/V_{rel}^2$$

Then, the time evolution of their mass ratio is

$$d/dt \, (M_1/M_2) = M_1/M_2 \, (1/M_1 \, dM_1/dt - 1/M_2 \, dM_2/dt) > 0.$$

This mean that the mass ratio *increases* with time. In other words, initially small differences in mass among the planetesimals are rapidly magnified, in an exponential manner. This growth mode is called *runaway growth* (Greenberg al., 1978; Wetherill and Stewart, 1989).

It is evident from this description that runaway growth occurs as long as there are objects in the disk for which $V_{esc} \gg V_{rel}$. The velocity dispersion of the planetesimals, however, is influenced by the escape velocity from the largest bodies in the disk. A planetesimal that experiences a near-miss with the largest body has its trajectory permanently perturbed and will have a relative velocity $V_{rel} \sim V_{esc}$ upon the next return. Thus, the planetesimals tend to acquire relative velocities of the order of the escape velocity from the most massive bodies, and when this happens runaway growth is shut off (see below).

From the discussion above, the conditions for runaway growth, when they are met, appear to hold only temporarily. There are, however, dynamical damping effects that may help keeping them valid for longer times. The first effect is that of gas drag. Gas drag not only causes the drift of bodies towards the central star, as seen above, but it also tends to circularize the orbits, thus reducing their relative velocities ($V_{rel}$). Whereas orbital drift vanishes for planetesimals larger than about 1 km in size, eccentricity damping continues to influence bodies up to several tens of kilometres across. Notice, however, that in a turbulent disk gas drag cannot damp $V_{rel}$ down to zero: in presence of turbulence the relative velocity evolves towards a size-dependent equilibrium value (Ida et al., 2008).

The second damping effect is that of collisions. Particles bouncing off each other tend to acquire parallel velocity vectors, reducing their relative velocity to zero. For a given total mass of the planetesimal population, this effect has a strong dependence on the planetesimal size, roughly $1/r^4$ (Wetherill and Stewart, 1993; Goldreich et al., 2004).

It is clear from this discussion that, in order to have an extended phase of runaway growth in a planetesimal disk, it is essential that the bulk of the solid mass is in small planetesimals, so that the damping effects are important. Because small planetesimals collide with each other very frequently and, upon collisions, comminute themselves into dust or grow by coagulation, this condition may not hold for a long time. Moreover, if planetesimals really form with a preferential size of ~100km, as in the streaming instability scenario, the population of small planetesimals would have been insignificant and therefore runaway growth would have lasted only very shortly, if ever.

**ii) Oligarchic growth**

When the dispersion velocity of the planetesimals $V_{rel}$ becomes of the order of $V_{esc(1)}$, the gravitational focussing factor $F_g$ becomes of order unity. Consequently the mass growth equation becomes

$$1/M_1 \, dM_1/dt \sim M_1^{-1/3}$$

that is, the relative growth rate of the large bodies gets slower and slower as the bodies grow. The time evolution of the mass ratio between two bodies of masses $M_1$ and $M_2$ with $M_1 > M_2$ becomes:

$$d/dt \, (M_1/M_2) = M_1/M_2 \, (1/M_1 \, dM_1/dt - 1/M_2 \, dM_2/dt) < 0,$$

so that the mass ratios among the large bodies tend to converge to unity.

In principle, one could expect that the small bodies also narrow down their mass difference with the large bodies. But in reality, the large value of $V_{rel}$ prevents the small bodies from accreting one another. The small bodies can only contribute to the growth of the large bodies (i.e. those whose escape velocity is of the order of $V_{rel}$). This phase is called *oligarchic growth* (Kokubo and Ida, 1998, 2000).

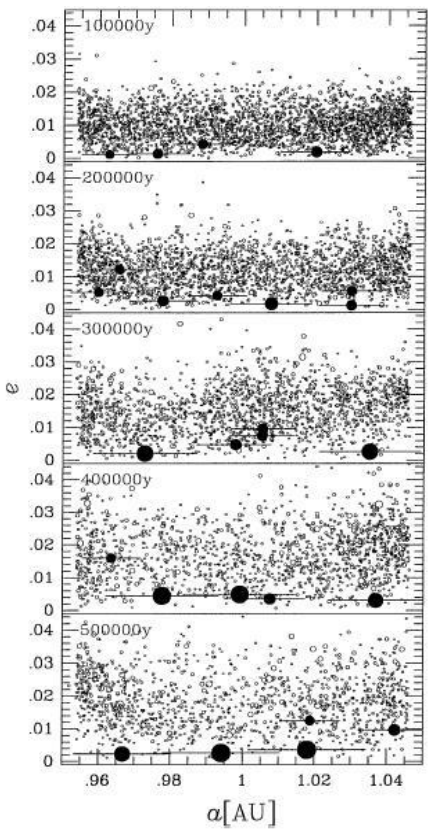

Fig. 5. Snapshots of a planetesimal system, illustrated on the semi-major axis vs. eccentricity plane. The circles represent the planetesimals and their radii are proportional to the physical radii. The filled circles represent oligarchs with masses larger than $2 \times 10^{25}$g. From Kokubo and Ida, 2000.

In practice, as shown in Fig. 5, oligarchic growth leads to the formation of a group of objects of roughly equal masses, embedded in the disk of planetesimals. The mass gap between oligarchs and planetesimals is typically of a few orders of magnitude. Because of dynamical friction, planetesimals have orbits that are much more eccentric than the oligarchs. The orbital separation among the oligarchs is of the order of 5 to 10 mutual Hill radii $R_H$, where:

$$R_H = (a_1+a_2)/2 \, [(M_1+M_2)/3M_S]^{1/3}$$

$a_1$ and $a_2$ are the semi-major axes of the orbits of the objects with masses $M_1$ and $M_2$, and $M_S$ is the mass of the star.

**Problems**

The classic view of planet formations is that the processes of runaway growth and oligarchic growth convert most of the planetesimals mass into a few massive objects: the protoplanets. This view, however, does not pass a closer scrutiny.

In the Solar system, two categories of proto-planets formed within the lifetime of the gas component of the protoplanetary disk (i.e. a few millions of years, according to observations of protoplanetary disks around young stars, Haisch et al., 2001). In the outer system, a few planets of multiple Earth masses formed, i.e. massive enough to be able to capture a substantial mass of H and He from the disk and become the observed giant planets, from Jupiter to Neptune. In the inner disk, instead, the proto-planets had only a mass of the order of the mass of Mars which eventually formed the terrestrial planets after the disappearance of the gas (see below for the formation of giant and terrestrial planets). Thus, the proptoplanets in the outer part of the disk were 10-100 times more massive of those in the inner disk. This huge mass ratio is even more surprising if one considers that the orbital periods, which set the natural clock for all dynamical processes including accretion, are ten times longer in the outer disk.

A natural divide between the inner and the outer disk, is the snowline. The surface density of solid material is expected to increase through the snowline, due to the availability of water ice. However, a recent evaluation of the oxygen abundance in the Sun (Lodders, 2003) shows that this density-increase is only of a factor of ~2. This is insufficient to explain the huge mass ratio between proto-planets in the outer and inner parts of the disc (Morbidelli et al., 2015).

In addition, whereas in the inner disc the process of oligarchic growth can continue until most of the planetesimals have been accreted by proto-planets, the situation is much less favourable in the outer disc. There, when the protoplanets become sufficiently massive (about 1 Earth mass), they tend to scatter the planetesimals away, rather than accrete them. In doing this they clear their neighbouring region, which in turn limits their own growth (Levison et al., 2010). In fact, scattering dominates over growth when the ratio $v_{esc}^2/2v_{orb}^2 > 1$, where $v_{esc}$ is the escape velocity from the surface of the protoplanet and $v_{orb}$ is its orbital speed (so that $2\,v_{orb}$ is the escape velocity from the stellar potential well from the orbit of the protoplanet). This ratio is of course much larger in the outer disc than in the inner disc because $v_{orb}^2 \sim 1/a$, where $a$ is the orbital semi major axis.

Therefore, understanding the formation of the multi-Earth-mass cores of the giant planets and their huge mass ratio with the protoplanets in the inner Solar System is a major problem of the runaway/oligarchic growth models, and it has prompted the elaboration of a new planet growth paradigm, named *pebble accretion*.

**Pebble accretion**

Let's take a step back to what seems to be most promising planetesimal formation model: that of self-gravitating clumps of small particles (hereafter called *pebbles* even though in the inner disc they are expected to be at most mm-size, so that *grains* would be a more appropriate term). Once a planetesimal is formed, it remains embedded in the disc of gas and pebbles and therefore it can keep growing by accreting individual pebbles. This process has been first envisioned by Ormel and Klahr (2010) and then studied in details in Lambrechts and Johansen (2012, 2014).

Pebble accretion is much more efficient than planetesimal accretion because of two reasons. First, the accretion cross-section for a planetesimal-pebble encounter is much larger than for a planetesimal-planetesimal encounter. As seen above, in a planetesimal-planetesimal encounter the accretion cross-section is $\pi r^2 F_g$, where r is the physical size of the planetesimal and $F_g$ is the gravitational focussing factor. But in a planetesimal-pebble encounter it can be as large as $\pi R_g^2$, where $R_g$ is the distance at which the planetesimal can start bending the trajectories of the incoming objects. This is because, as soon as the pebble's trajectory starts to be deflected, its relative velocity with the gas increases and gas-drag becomes very strong. Thus, the pebble's trajectory spirals towards the planetesimal. This is shown in the inlet of Fig. 6. The gravitational deflection radius $R_g$ is either the Bondi radius $R_B=GM/v_{rel}^2$ or the Hill radius $R_H=a(M/3M_S)^{1/3}$, whichever is smaller, and it is typically much larger than the physical radius of the planetesimal. The outer panel of Fig. 6 shows the actual accretion radius as a function of the pebble's friction time.

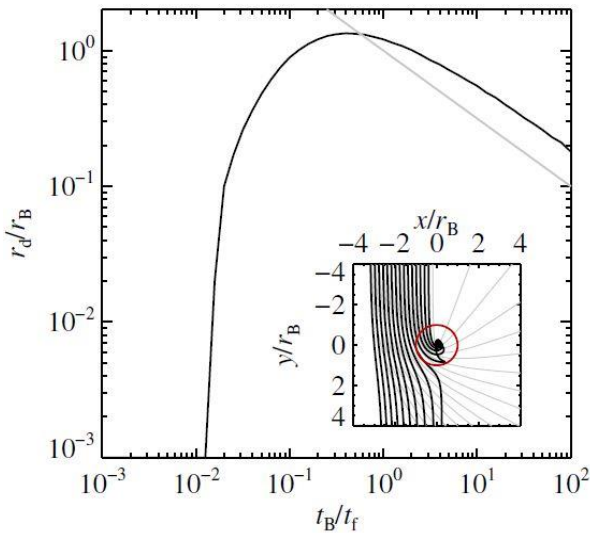

Fig. 6. The outer plot shows the accretion radius $r_d$, normalized to the Bondi radius $R_B$, as a function of $t_B/t_f$, where $t_f$ is the friction time and $t_B$ is the time required to cross the Bondi radius at the encounter velocity $v_{rel}$. The smaller is the pebble the larger is $t_B/t_f$. The inset shows pebble trajectories (black curves) with $t_B/t_f =1$, which can be compared with those of objects with $t_B/t_f \to 0$ (grey curves). Clearly the accretion radius for the former is much larger. A circle of Bondi radius is plotted in red. From Lambrechts and Johansen, 2012.

The second reason for which pebble accretion is more efficient than the accretion of other planetesimals is that pebbles drift in the disk. Thus, the orbital neighborhood of the growing body cannot become empty. Even if the growing body accretes all the pebbles in its vicinity, the local population of pebbles will be renewed by new particles drifting from larger distances. This cannot happen for planetesimals, given that their radial drift in the disc is negligible.

Provided that the mass-flux of pebbles through the disk is large enough, pebble accretion has been shown to be able to grow the largest planetesimals up to multiple Earth-masses, i.e. to form the giant planets cores within the disc's lifetime (Lambrechts and Johansen, 2012, 2014). The large mass ratio between protoplanets in the outer vs. inner parts of the disc can be explained by remembering that icy pebbles can be relatively large (a few centimeters in size), whereas in the inner disc the pebble's size is limited to sub-millimetre by the bouncing silicate barrier (chondrule-size particles) and by taking into account that pebble accretion is more efficient for large pebbles than for chondrule-size particles (Morbidelli et al., 2015).

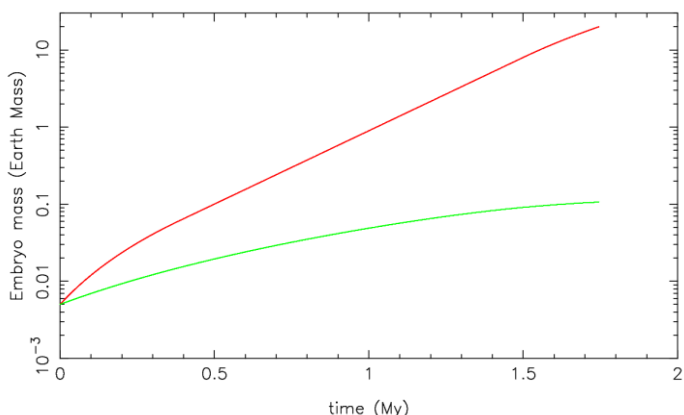

Fig. 7. The mass growth of two planetesimals in the outer disc (red) and inner disc (green). Intitially both planetesimals have ½ of the Lunar mass; the one in the outer disc accretes cm-size pebbles, whereas the one in the inner disc accretes mm-size pebbles, with ½ of the total mass flux (due to the sublimation of ice at the snowline). Thus, the planetesimal in the outer disc grows to 20 Earth masses in the time when the planetesimal in the inner disc reaches only one mass of Mars. (from Morbidelli et al., 2015).

For all these reasons, despite some unknown still subsist, pebble accretion is now considered to be the dominant process leading to planet formation.

## Different categories of planets, different accretion histories

There are, grossly speaking, three categories of planets. Giant planets have massive envelopes of H and He, with a total mass exceeding the solid mass of the planet. Jupiter and Saturn are the giant planets of the Solar System. Terrestrial planets are much smaller in mass and have no primordial H and He in their atmospheres, i.e. captured from the protoplanetary disk. Instead, their atmospheres, where they exist, are made of gases released from the interior of the planet. Super-Earths or Neptune-like planets (the acronym SEN, proposed by Mayor et al., 2011, will be adopted here) are an intermediate category. More massive than terrestrial planets but less massive than giant planets, they have atmospheres of H and He captured from the disc, but the total mass of these gases is smaller than the solid mass of the planet. The growth histories of the planets in these three categories are, of course, different, as described below, but can be understood in the common framework of accretion outlined above.

**Giant planets.**

Planets accreting enough pebbles can grow to multiple (even tens) of Earth masses. Because of their relevant gravitational pull, they tend to capture in their gravitational well also some gas from the disc. As long as the accretion of solid particles is vigorous, only a small amount of gas can be captured. This is because the potential energy released by the accretion of solids heats the gas, and the thermal velocity of the gas molecules exceeds the escape velocity from the planet. However, when the planet achieves a large enough mass (about 20 Earth masses in the outer disc) pebble accretion suddenly stops. This is because the gravitational effects of the planet alter the disc's gas distribution in the vicinity of the planet's orbit. If these effects are sufficiently strong, an overdense ring of gas can form just beyond the planet's orbit, with orbital velocity exceeding the Keplerian speed. The super-Keplerian rotation of the gas in this ring reverses the drag on the pebbles. Instead of being slowed-down by the headwind, pebbles are now accelerated by a tail-wind. They don't spiral towards the central star, but spiral outwards. Thus, the flux of pebbles towards the planet is interrupted and the planet's accretion of solid stops (Lambrechts et al., 2014). Without the heat released by solid accretion, gas can now be more effectively bound to the planet. When the mass of the gas envelope exceeds the solid mass of the planet, the accretion accelerates exponentially because of a positive feedback: the more gas is accreted, the more massive becomes the planet and therefore the larger becomes the planet's ability to accrete more gas. The accretion rate of gas is limited by the rate of radial drift of gas through the disc, which depends on the effectiveness of its angular momentum redistribution. Eventually, gas accretion stops when the gas is removed from the disc by photoevaporation processes.

It is important to mention that, because of the gravitational interactions between the planet and the disc of gas, the planet cannot remain on its original orbit. Planet migration is a ubiquitous process and dictates the final location of the planets in the system. As long as the planet has a moderate mass (few tens of Earth masses) the migration speed is proportional to the planet's mass (Type-I migration; Tanaka et al., 2002). Type-I migration is typically directed towards the star, but it can be halted under some conditions, for instance in a region of sharp radial increase of the gas density in the disc (i.e. at the inner disc's edge or at the outer edge of a partial cavity; Masset et al., 2006; Paardekooper et al., 2010). When the planet exceeds tens of Earth masses, it starts carving a gap along its orbit in the disc's gas distribution. This affects the migration process. The planet now has to move together with its gap, and therefore it has to follow the redistribution of gas in the disc (Type-II migration; Lin and Papaloizou, 1986). The latter depends on the ability of gas to exchange angular momentum, namely on the gas viscosity. The actual viscosity of discs is unknown, but it is expected to be small. Therefore, Type-II migration of massive (giant) planets is expected to be slower than the Type-I migration of planets of a few Earth masses.

Most extrasolar giant planets discovered to date are at about 1-2 AU from the central star. Supposing that these planets formed quickly at the snowline location of their respective disks (say at distances of 5, possibly 10 AU), this observation suggests that Type-II migration is slow enough to move giant planets by only a few AUs within the lifetime of the disk. Jupiter and Saturn are beyond 5 AU today. It is not known if this is an oddity relative to the real distribution of extrasolar giant planets because the bias against the detection of planets by the radial velocity technique becomes too severe beyond a few AUs. However, extrasolar planet surveys by direct imaging show that planets beyond tens of AUs are very rare. It has been shown (Masset and Snellgrove, 2001) that the gravitational interaction between Jupiter and Saturn in the disc can stop and even reverse their Type-II migration. This opens the intriguing possibility that Jupiter was once much closer to the Sun, at an orbital distance comparable to that of the observed extrasolar giant planets, and then migrated outwards under the influence of Saturn until it reached the current distance at the end of the disc's lifetime (Walsh et al., 2011).

**Terrestrial planets**

Radioactive chronometers applied to terrestrial mantle rocks show that the Earth took tens of millions of years to form (Halliday and Kleine, 2006). This timescale is much longer than the typical lifetime of a protoplanetary disc (Haisch et al., 2001). Mars, instead, formed very quickly, in a few My, i.e. within the disc's age (Dauphas and Pourmand, 2011). It is also known that the Moon formed as consequence of an impact of a Mars-mass body with the Earth (strictly speaking this is a model, but it is the only one that explains the observations; see Canup, 2014 for a review). Giant impacts have probably been ubiquitous in terrestrial planet formation. The high iron/silicate fraction in Mercury suggests that this planet has lost a substantial fraction of its mantle in one (or more) giant impacts. If one restores its supposed original chondritic composition, Mercury also had a mass comparable to that of Mars. Taken together, all these elements strongly suggest that, as already anticipated above, numerous Mars-mass protoplanets formed in the inner part of the Solar

System's disc. Because of their small masses these protoplanets could not capture significant envelopes of H and He. Moreover, they did not migrate significantly in the disc, because the migration speed is proportional to the proto-planet's mass and, for a Martian mass, it is quite slow.

Numerical simulations show that, once the gas disappears from the disc, this system of multiple Mars-mass protoplanets naturally becomes unstable. The orbits of the protoplanets become eccentric, and start to intersect each other. This triggers a phase of giant impacts among the protoplanets. Through giant impacts, some planets (i.e. the Earth and Venus) grow to larger masses, on a timescale of tens to 100 My (Chambers and Wetherill, 1998; Chambers 2001; O'Brien et al., 2006). This process explains quite well all the available observational constraints on the formation of the terrestrial planets but one: the small mass of Mars. In fact, according to the simulations, also Mars should have grown up to almost an Earth mass. To prevent the further accretion of Mars beyond the disc's lifetime, it is necessary that the region beyond 1 AU was strongly depleted in solids (Hansen, 2009). This could have been the consequence of the previous migration history of Jupiter (Walsh et al., 2011) or of the possibility that the streaming instability operated in the early disc only within 1 AU (Drazkowska et al., 2016), so that planetesimals could grow to protoplanets only in that region.

**Super-Earths / Neptune-like planets (SENs)**

These planets are intermediate between those of the two categories above, not only because they are more massive than the Earth and less massive than the gaseous giant planets, but also for physical structure and formation history. An important difference with the terrestrial planets is that SENs achieved most of their final mass within the lifetime of the protoplanetary disc of gas. Consequently, orbital migration played an important role in setting their final orbits. Indeed, SENs are abundant on short-period orbits, typically with semi major axes well smaller than that of the orbit of Mercury. Also, by achieving a large mass before the disappearance of the gas, the SENs could accrete an atmosphere of H and He. However, the mass of the accreted gas did not overcome the mass of the solid planet, so that runaway gas accretion never started. This is why SENs did not evolve to the status of gas giant planets. Notice moreover that the SENs on very short period orbits, strongly irradiated by their parent star, might have lost their primordial atmosphere after the removal of the proto-planetary disc.

The Solar System has two SENs: Uranus and Neptune. The major difference with the SENs observed around other stars is that Uranus and Neptune remained on distant orbits, instead of migrating to the close proximity of the parent star. It is believed that this is because Jupiter and Saturn formed closer to the Sun than Uranus and Neptune and they set into an orbital configuration that prevented inward migration (Masset and Snellgrove, 2001). Thus, Uranus and Neptune could not be free to migrate towards the Sun, but were retained beyond the orbits of the gas-giants, in orbital resonances (i.e. on orbits whose periods are in integer ratios with those of Jupiter and Saturn; Morbidelli et al., 2007). Actually, the presence of Jupiter and Saturn may have favoured the formation of Uranus and Neptune. In fact, the large obliquities of Uranus and Neptune suggest that these planets did not form solely by pebble accretion, but witnessed a phase of giant impacts, like the terrestrial planets. Izidoro et al. (2015) showed that a system of multiple SENs, with a mass of the order of 5 Earth masses, whose migration is inhibited by the presence of Jupiter and Saturn, evolves in a natural way into a system of 1-3 Neptune-mass planets through a sequence of mutual collisions. Izidoro et al. (2017) then showed that, in absence of Jupiter and Saturn, these SENs would have instead migrated to the inner edge of the protoplanetary disc. The very compact SEN system produced by the migration process is likely to become unstable after the disappearance of the disc. The final orbital distributions of the planets of systems that experienced such instability reproduce very well the orbital characteristics of the observed extrasolar SENs.

# And now what?

Extrasolar planet searches have shown that planet formation is a ubiquitous process around stars. Despite solid accretion is so easy for Nature, it is still hard to understand for human scientists. This is because of the wide range of spatial and temporal scales involved, which makes difficult to enact the accretion process in laboratory experiments or in virtual computer simulations.

After decades of research, a new paradigm is emerging in which grain/pebble-size particles play the key role. In this paradigm, planetesimals would have been generated from self-gravitating clusters of these particles, which had clumped together due to their interaction with the (possibly turbulent) gas of the proto-planetary disk. Then, once formed, the largest planetesimals would have continued to accrete grains and pebbles individually but in large number, growing spectacularly in mass until becoming (proto)planets.

However, despite this general picture appears satisfactory, areas of unknown persist. The formation of planetesimals by the particle-clustering mechanism requires either large particles or a large solid/gas ratio. But particles were probably not larger than a millimetre in size, particularly in the inner part of the disk where chondrules have been the building blocks of planetesimals. And the solid/gas ratio in the Solar System was only ~ 1%. So, understanding the conditions of formation of planetesimals is not obvious. It is possible that radial migration of grains/pebbles, increased the solid/gas ratio locally, at specific locations in the disk. This process could have been at the origin of the first planetesimals, those that eventually differentiated (some of which are parent bodies of iron meteorites). Another possibility is that the solid/gas ratio increased when the gas was undergoing photo-evaporation from the central star, forming planetesimals near the end of the gas-disc lifetime. However, this requires that the solids remained somehow preserved in the disc, despite their tendency to drift towards the star by gas drag.

As for the formation of planets, the pebble-accretion process is effective if there are quite massive accretion seeds. Asteroid-size planetesimals don't accrete pebbles efficiently enough to become planets within a few My. Objects with masses comparable to the mass of the Moon are needed to trigger planet-growth. The origin of these quite-massive objects is unclear. Did they form directly from exceptionally massive self-gravitating clumps of particles? Or did they form from the collisional coagulation of planetesimals? There is currently no clear answer to this question.

The efforts of the scientific community will focus on these open issues. Besides the theoretical effort, observations of protoplanetary discs may provide crucial information on particle growth, particle drift and redistribution in the disc, and origin of protoplanets in discs. The evidence for a generic ring-like structure of protoplanetary discs provided by ALMA already shows that the view of continuous coagulation and migration of grains is probably simplistic. On a different front, cosmochemistry is entering a new era in which measurements of isotopic ratios with exquisite precision become possible. This allows scientists to find the trace of distinct isotopic reservoirs in the Solar System disc, re-asses chemical and isotopic complementarities among the constituents of meteorites, determine precisely absolute formation ages. All these data will provide crucial clues to unveil how planetesimals and planet formation occurred, at least in the Solar System.

Once massive protoplanets grow in the disc, a complex interplay starts between accretion and migration. This complexity is probably the key to understand the great diversity among planetary systems. Few general tendencies begin to appear in simulations. If pebble-accretion is inefficient, the proto-planets remain small and therefore they don't migrate significantly through the disc. After removal of the gas, these small and numerous protoplanets can become unstable and, through a series of giant impacts, can form moderate-mass planets such as the terrestrial ones. If pebble accretion is more efficient, fewer and bigger protoplanets form, and they migrate towards the inner edge of the disc, giving origin to systems of close-in superEarths. And if pebble-accretion is even more efficient, some protoplanets grow enough to capture large masses of hydrogen and helium from the disc and become gas-giant planets. The migration of these planets is much slower and their presence affects the growth and the migration of all the other planets in the system. The specific structure of the Solar System can be understood from the early growth of Jupiter and Saturn. The full complexity of the interplay between accretion and migration, however, remains to be explored.

Planetary science is living exciting times, in which the synergy between astronomical observations, geochemical and cosmochemical analyses, in-situ exploration of the objects of the Solar System and theoretical modelling are leading to a leap forward in understanding of planet formation that has no precedent in this field.